\documentclass[]{spie}  

\usepackage{amsmath,amsfonts,amssymb}
\usepackage{graphicx}
\usepackage[colorlinks=true, allcolors=blue]{hyperref}
\usepackage{orcidlink}
\usepackage{subfigure}

\title{A Focal-plane X-ray Polarimeter with Spectral and Timing Capabilities for Future Missions}

\author[a]{Hemanth Manikantan\orcidlink{0000-0001-9404-1601}}
\author[a]{Carlo Lefevre}
\author[a]{Lorenzo Petrucci\orcidlink{0009-0007-7331-3680}}
\author[a]{Paolo Soffitta\orcidlink{0000-0002-7781-4104}}
\author[a]{Fabio Muleri\orcidlink{0000-0003-3331-3794}}
\author[a]{Enrico Costa\orcidlink{0000-0003-4925-8523}}
\author[a]{Alda Rubini}
\author[b]{Vladislavs Plesanovs\orcidlink{0000-0003-3919-3068}}
\author[b]{Markus Gruber\orcidlink{0000-0002-4767-1392}}
\author[b]{Jochen Kaminski\orcidlink{0000-0001-5979-6996}}
\author[b]{Klaus Desch\orcidlink{0000-0001-5836-6118}}
\author[a]{Alessandro Di Marco\orcidlink{0000-0002-1159-4460}}
\author[a]{Sergio Fabiani\orcidlink{0000-0003-1533-0283}}
\author[a]{Riccardo Ferrazzoli\orcidlink{0000-0003-1074-8605}}
\author[a,c]{Saba Imtiaz\orcidlink{0009-0001-1972-7774}}
\author[a]{Dawoon E. Kim\orcidlink{0000-0001-5717-3736}}
\author[a]{Alessandro Lacerenza}
\author[a,d]{John Rankin\orcidlink{Rankin 0000-0002-9774-0560}}
\author[a,e]{Ajay Ratheesh\orcidlink{0000-0003-0411-4243}}
\affil[a]{INAF Istituto di Astrofisica e Planetologia Spaziali, Via del Fosso del Cavaliere 100, Rome, Italy}
\affil[b]{Physikalisches Institut, Universität Bonn, Nussallee 12, Bonn, Germany}
\affil[c]{Padua University, Via Marzolo 8, Padua, Italy}
\affil[d]{Dipartimento di fisica e scienze della terra, Università di Ferrara, Via Giuseppe Saragat, 1, Ferrara, Italy I-44122}
\affil[e]{Technical University of Denmark, Elektrovej 327, 2800 Kgs. Lyngby, Denmark}

\authorinfo{Further author information: (Send correspondence to H.M.)\\E-mail: hemanth.manikantan@inaf.it}

\begin{document} 
\maketitle

\begin{abstract}
The NASA-ASI mission \textit{Imaging X-ray Polarimeter Explorer} (\textit{IXPE}) firmly established X-ray polarimetry as a core observation pillar of high-energy astrophysics, alongside imaging, timing, and spectroscopy. While \textit{IXPE} made groundbreaking discoveries in the 2--8 keV range on both point and extended X-ray sources, the \textit{Gas Pixel Detectors} onboard \textit{IXPE} demand substantial improvements to be ready for the next generation of imaging X-ray polarimeters. Here, we present the development of a detector prototype that can deliver the next generation of sensitive imaging X-ray polarimetry across the energy band of 2--30 keV. We use Ar/DME based gas volume to absorb X-rays by the photoelectric effect and a pixelated CMOS readout ASIC, Timepix3, with excellent timing capabilities to image the resulting photoelectron track in three dimensions. The Timepix3 is integrated with a multiplication stage called InGrid that enables single-primary-electron detection from the gas volume. The photoelectron track reconstructed in three dimensions increases the polarimetric sensitivity towards the lowest operable energies, and the deadtime-free operation of the ASIC facilitates the usage as a focal plane instrument on high-throughput X-ray mirrors. With this prototype, we demonstrate a low-medium-energy X-ray polarimeter with excellent timing and moderate spectral capabilities.
\end{abstract}


\section{INTRODUCTION}
\label{sec:intro}  

X-ray polarimetry offers unique diagnostics of the magnetic field structure, particle acceleration mechanism, and scattering environment geometry in high-energy astrophysical sources. Historically, only a handful of significant polarimetric measurements were obtained in X-rays, like the polarization of the Crab Nebula with the Bragg polarimeter onboard the \textit{8$^\mathrm{th}$ Orbiting Solar Observatory} (\textit{OSO-8}) \cite{weisskopf1976measurement}. The \textit{Imaging X-ray Polarimeter Explorer} (\textit{IXPE}) is a NASA Astrophysics Small Explorer Mission (SMEX), which is the first X-ray astronomy mission dedicated to polarimetry, and was launched in December 2021 \cite{ixpe_weisskopf}. \textit{IXPE} demonstrated the broader scientific potential of imaging and spectral-polarimetry, enabling groundbreaking studies of a variety of sources, including extended sources like pulsar wind nebulae and supernova remnants, as well as point sources like magnetars, accreting neutron stars and black holes, and active galactic nuclei in the 2--8 keV band \cite{Soffitta:2024pmr}. The remarkable discoveries of highly uniform magnetic fields in the particle acceleration sites of the Vela pulsar wind nebula \cite{xie2022vela} and the Crab nebula \cite{bucciantini2023simultaneous} and polarized emission from the galactic center near Sgr A* \cite{marin2023x} underscore the potential of imaging polarimetry. 

\textit{IXPE} achieved imaging polarimetry through three \textit{Gas Pixel Detector} (GPD) units \cite{soffitta2021instrument,baldini2021design} placed at the focal plane of three respective Wolter-I mirror modules \cite{ramsey2022optics}. GPD operates with the principle of photoelectric track imaging \cite{costa2001efficient}. It uses a gas absorption volume (filled with Dimethyl ether (CH$_3$OCH$_3$) at 800 mBar) and a Gas Electron Multiplier (GEM) multiplication stage paired to a finely pixelated (50 $\mu$m pitch) CMOS readout ASIC XPOL \cite{bellazini_sealedgas,xpol_bellazzini,baldini2021design}. The X-ray photon passes through a beryllium window into the gas volume 1 cm deep (called the drift gap). The photon interacts with an atom in the gas volume by the photoelectric effect, which is the dominant interaction mechanism in this energy range, liberating a photoelectron, which subsequently leaves a trail of electron-ion pairs along its path by ionization. The photoelectron is ejected preferentially in the direction of photon polarization, and therefore the polarization information is contained in the initial part of the track. By applying an electric field across the drift gap, the electrons in the track are drifted towards the readout ASIC, very close to which a GEM stage allows for gas multiplication. The track extends over a few 100 $\mu$m, depending on the energy of the photoelectron. The pixel coordinates of collected electrons and their number constitute a measured photoelectron track, in which a statistical moments-based algorithm \cite{bellazini_moment} is employed to compute the photon impact position and the polarization direction. In the case of a preferential polarization angle ($\phi_\mathrm{P}$) in the absorbed X-ray photons, it manifests as a cos $2$($\phi$-$\phi_\mathrm{P}$) modulation in the distribution of measured azimuthal angles (called modulation curve).

Notwithstanding the remarkable achievements of \textit{IXPE}, the 2--8 keV GPD faces several technological limitations. These include restricted count-rate capabilities for bright sources caused by ASIC dead time, and challenges in rejecting charged-particle background. Due to the 1 ms dead-time limitation of GPD, observing very bright sources is not feasible, and it has to use onboard filters to reduce the count rate of sources that are twice brighter than the Crab \cite{soffitta2021instrument}. For the same reason, pairing these detectors with the next generation of high-throughput X-ray optics (e.g., \cite{reynolds2024advanced}) is not possible. The high count rate operability requires an ASIC capable of recording events with dead time much less than 1 ms. The difficulty in background rejection stems from incomplete track imaging, a consequence of the relatively low gas gain ($\sim$100) \cite{baldini2021design, dimarco_background}. 

Beyond electronics constraints, polarization sensitivity at the lower end of the energy band is fundamentally limited by track morphology. The polarization sensitivity for a detector is quantified as the Minimum Detectable Polarization at 99\% confidence level. (MDP$_{99}$), which is the minimum degree of polarization that can be distinguished from noise at 99\% c.l.. MDP$_{99}$ is inversely proportional to the \textit{Quality Factor}  (\textit{QF} = $\mu \sqrt{\epsilon}$), where $\mu$ is the modulation factor measured for the instrument for a 100\% polarized source and $\epsilon$ is the quantum efficiency of the detector. Due to the design of the gas X-ray polarimeters, a compromise between $\mu$ and $\epsilon$ must be achieved, as an improvement in one parameter inherently degrades the other. While $\epsilon$ can be increased by using higher-$Z$ gas, increasing gas pressure, or utilizing a thicker drift region, these modifications would deteriorate the track sizes and their diffusion during drift, which ultimately reduces $\mu$. Three-dimensional reconstruction of photoelectron tracks has been proposed as an alternate means to enhance $\mu$ at the lower end of the operational energy band without compromising $\epsilon$ \cite{kim2024future}. The primary requirement for achieving this 3D track reconstruction is excellent time resolution (of the order of ns) to capture the relative vertical distance between track segments.

A promising solution to achieve 3D track reconstruction, overcome dead time limitations, and enable proper charged particle imaging is offered by the hybrid gas detector GridPix, which combines a Timepix3 pixel ASIC with an integrated Micromegas (InGrid) gas-amplification stage coupled to a photon absorption gas volume \cite{gruber2025development}. Timepix3 is a general-purpose pixel readout ASIC manufactured in a 130 nm CMOS process. The ASIC incorporates 256$\times$256 square pixels with a pitch of 55 $\mu$m, each with an independently adjustable threshold. In the \emph{Simultaneous TOA and TOT} mode, each pixel simultaneously measures the signal threshold crossing time (TOA) and the time over threshold (TOT), which is proportional to the deposited charge in the pixel. TOA is measured with a global 40 MHz clock, providing a coarse timing resolution of 25 ns. For improved timing precision, an optional gated 640 MHz clock is enabled to record a fine TOA (fTOA), yielding an effective timing resolution of 1.5625 ns. TOT is measured in cycles of the 40 MHz clock. The intrinsic per-pixel dead time is 475 ns plus the measured TOT. The chip features a zero-suppressed data-driven readout architecture in which only the pixels that cross the threshold are transmitted, allowing dead-time-free operations up to hit rates of 40 Mhits s$^{-1}$ cm$^{-2}$ \cite{poikela2014timepix3}. These enable reconstruction of photoelectron tracks in three dimensions and operability with bright sources \cite{fabiani2024towards} as well as X-ray optics with large effective area \cite{reynolds2024advanced}. The improved gas gain (few $\sim$1000) in the InGrid enables detection of minimum ionizing particle tracks, which otherwise appear as multiple short tracks in \textit{IXPE}/GPD. This allows distinguishing the track morphology and effectively allows better background rejection. 

The GridPix detectors were conceived for Solar Axion-like particle search, and were used for the CAST experiment and will be employed in the future for Baby-IAXO\cite{krieger2017gridpix, schiffer2025detector}. To utilize this detector for performing photoelectric polarimetry, they have to be tuned. This tuning is not only limited to the gas mixture, but also to the detector geometry, to make a uniform drift field. This is because the non-uniform drift field will distort the tracks during drift and result in distorted tracks imaged by Timepix3, subsequently impacting the reconstruction of the actual photoelectron ejection direction.

From the scientific perspective, expanding polarimetry to higher energies presents significant opportunities to probe the previously unexplored domains \cite{soffitta2024considerations}. Up to about 30 keV, photoelectric polarimetry remains a highly effective technique for imaging X-ray polarimetry. The core requirements for extending photoelectric polarimetry up to the 30 keV regime are a thicker drift gap and the use of noble gas mixtures based on higher-Z noble gases, such as argon or neon \cite{muleri2006x}.

In this article, we present the status of development and initial characterization of a GridPix-based detector optimized for photoelectric polarimetry.

\section{DETECTOR CONFIGURATION}

\subsection{Gas Cell -- Drift and Multiplication}

The gas mixture, gas cell thickness, and gas pressure are selected based on the energy range of operation and a sweet spot for the QF. Gas mixtures with heavier elements like argon have been shown to improve the photodetector-polarimetric capability in $\sim$3--30 keV \cite{muleri2006x}. The lower limit in energy comes from unpolarized Auger electron emission (Ar K-edge is 3.2 keV), while the upper limit comes from the quantum efficiency.

The detector base consists of a printed circuit board (PCB) carrying the Timepix3/InGrid assembly, to which a Vespel layer with a copper anode (hereafter referred to as the anode) is bonded. A PEEK detector body and a titanium frame housing a beryllium window (hereafter, the window) are then mounted onto this base. This highly modular design allows PEEK bodies, titanium frames, and windows of varying thicknesses to be easily interchanged. The primary drift field is established between the anode and the window. To prevent field distortions and enable long drift lengths, the PEEK body incorporates guard ring electrodes between these two planes. By utilizing different combinations of these components, the instrument can achieve drift lengths of 1--3 cm and sustain gas pressures of 1--3 bar, enabling polarimetric measurements across the 2--30 keV energy band \cite{muleri2006x}.

At the bottom of the drift cell lies the Timepix3 ASIC, equipped with an InGrid gas multiplication stage \cite{gruber2022development}. The InGrid is a perforated Aluminum mesh fabricated directly onto the Timepix3 via photolithographic techniques, ensuring the mesh holes are precisely aligned with the ASIC pixels. By applying a potential difference of $\sim$430 V across the gap between the grid and the Timepix3, an electric field of the order of $\sim$86,000 V cm$^{-1}$ is generated. This field yields a gas gain of $\sim$4000, meaning each primary electron arriving at a grid hole is amplified into an avalanche of approximately 4000 electrons before inducing a signal on the underlying pixel. The Timepix3 features per-pixel comparators for signal detection. The pixel thresholds were set to $\sim$900 electrons, achieving a single primary electron detection probability of $\sim$90\% at this operational gas gain.

\begin{figure}
    \centering
    \includegraphics[width=0.5\linewidth]{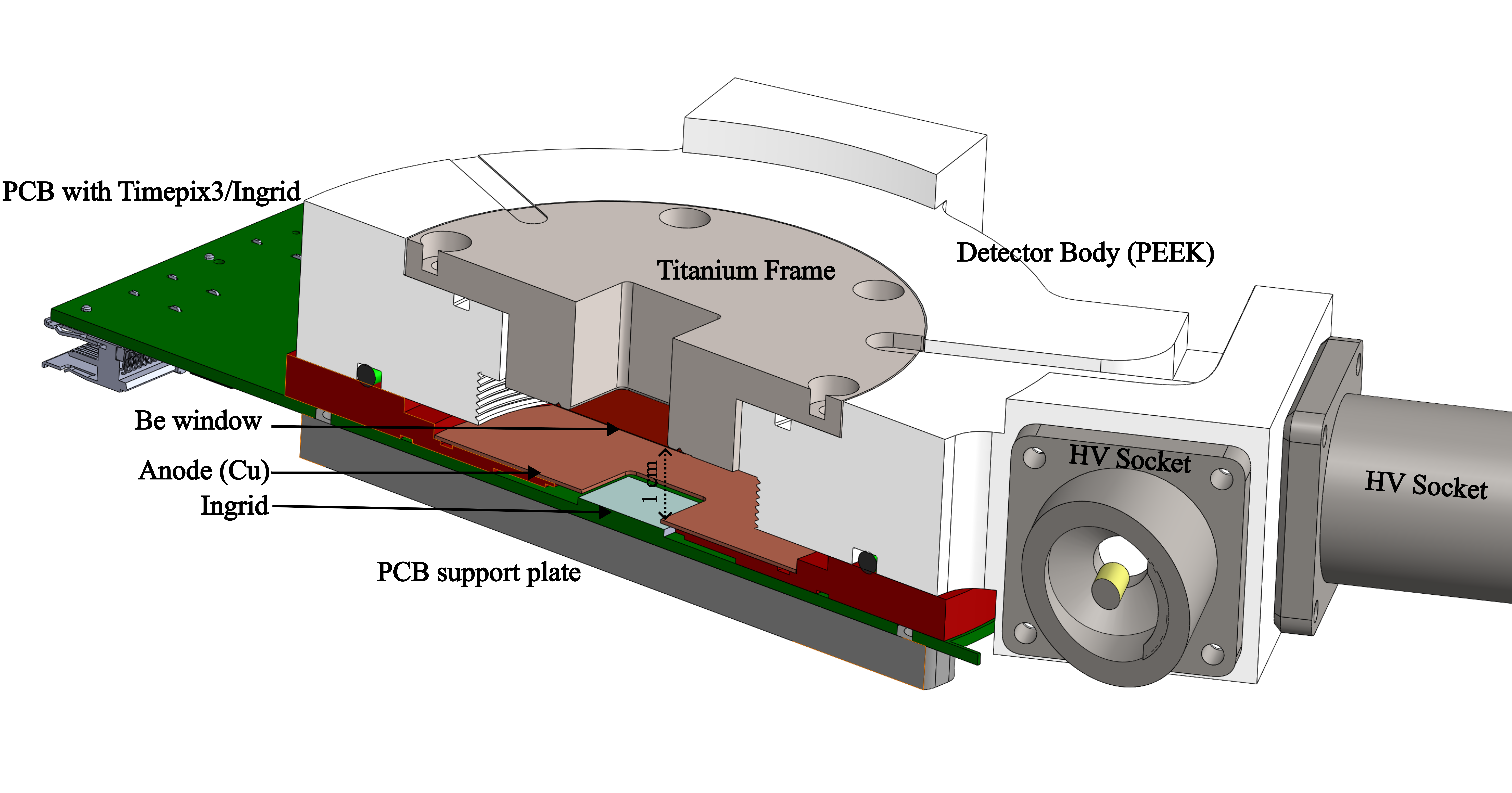}
    \includegraphics[width=0.27\linewidth]{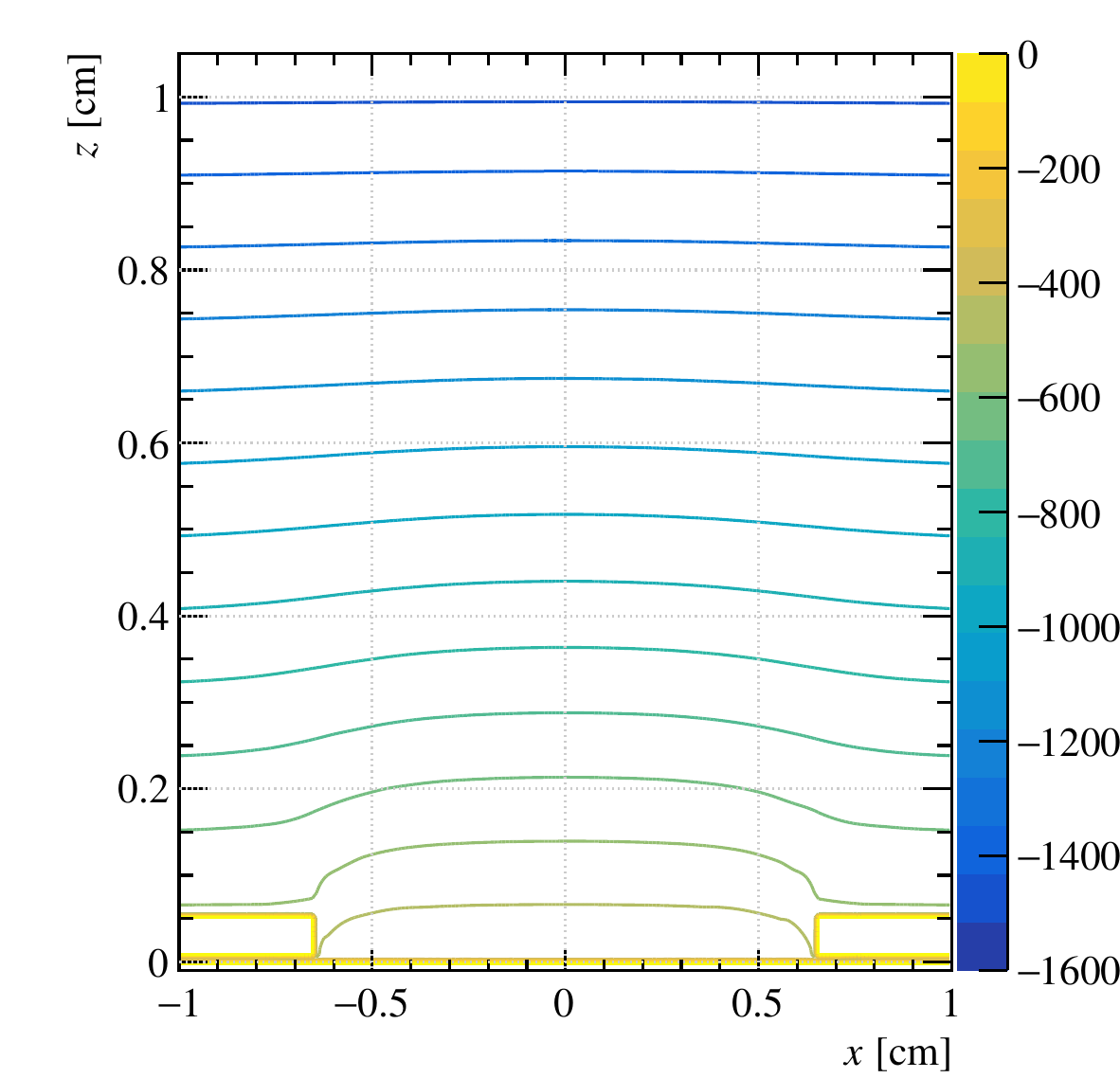}
    \caption{Left: Cross-section CAD view of the detector. Right: Equipotential lines in the 1 cm drift gap (viewed on the plane of cross section shown in CAD) when voltages of --1580, --580 and --430 V are applied to the Window (z=1 cm), Anode (z=0.009 cm) and Grid (z=0, -0.7$<$x$<$0.7 cm), respectively.}
    \label{fig:archetechture}
\end{figure}

In this work, we report the measurements with a 1 cm drift gap configuration with three electrodes (Figure~\ref{fig:archetechture}). We used a gas mixture of Ar/DME 80/20 at 1.2 bar (slightly above atmospheric pressure). The drift field of 1000 V/cm was selected based on minimum transverse diffusion (Figure~\ref{fig:diffusion_ArCO2}). This gas mixture has a W-value of 26.6 eV and quantum efficiencies of $\sim$31\% at 5.9 keV and $\sim$2\% at 17 keV.

\begin{table}[ht]
\caption{Specifications of the assembled detector.} 
\label{tab:detspec}
\begin{center}       
\begin{tabular}{l l} 
\hline
\hline
\rule[-1ex]{0pt}{3.5ex}  Gas Mixture & Ar/DME 80/20 1.2 bar  \\

\rule[-1ex]{0pt}{3.5ex}  Entrance window & Beryllium, 50 $\mu$m thick   \\

\rule[-1ex]{0pt}{3.5ex}  Detector body & PEEK, Vespel   \\

\rule[-1ex]{0pt}{3.5ex}  Readout ASIC & Timepix3/InGrid   \\

\rule[-1ex]{0pt}{3.5ex}  Pixel dimensions &Square $55\times55$ $\mu$m$^2$  \\

\rule[-1ex]{0pt}{3.5ex}  Threshold & $\sim900$ e-\\

\rule[-1ex]{0pt}{3.5ex}  Gas Gain & $\sim$4000\\

\rule[-1ex]{0pt}{3.5ex}  V$_\mathrm{ASIC}$ & 0 V  \\

\rule[-1ex]{0pt}{3.5ex}  V$_\mathrm{Grid}$ & --430 V  \\

\rule[-1ex]{0pt}{3.5ex}  V$_\mathrm{Anode}$ & --580 V\\

\rule[-1ex]{0pt}{3.5ex}  V$_\mathrm{Cathode}$ & --1580 V \\
\hline
\end{tabular}
\end{center}
\end{table} 

\subsection{Data Readout Architecture and Processing}
\label{sec:RA}

The readout system interfaces and transfers data between the Timepix3 ASIC and the control PC. It is built on the Scalable Readout System (SRS) developed by the RD51 collaboration \cite{toledo2011front}. The system consists of a power crate, a Front-end Concentrator (FEC) Card hosting a Virtex 6 FPGA, and a custom adapter PCB that connects the FEC to the Timepix3 carrier board. The firmware running on the FPGA is based on the Basil framework \footnote{\url{https://github.com/SiLab-Bonn/basil}, \url{https://github.com/GasDet-Bonn/tpx3-daq}} \texttt{v3.2.0} and communicates with the PC using the TCP protocol over a standard Ethernet cable. 

This architecture supports up to eight data links from Timepix3. To overcome the 14-bit counting limit of the internal 40 MHz clock on Timepix3, which would otherwise roll over every 409.6 $\mu$s, the FPGA periodically inserts a time-extension word to the data packets from Timepix3. This procedure, performed roughly every 100 $\mu$s, extends the effective timestamp range to several days, ensuring unambiguous time ordering during long measurements. Each 48-bit data packet from Timepix3 is transmitted to the PC as two separate 32-bit packets, which are then decoded by dedicated interpretation software. A complete description of the readout hardware, firmware, and data interpretation pipeline is present in \cite{gruber2022srs,gruber2025development}.

\begin{figure}[]
\centering
\subfigure[\label{fig:diffusion_ArCO2}]{\includegraphics[scale=0.68]{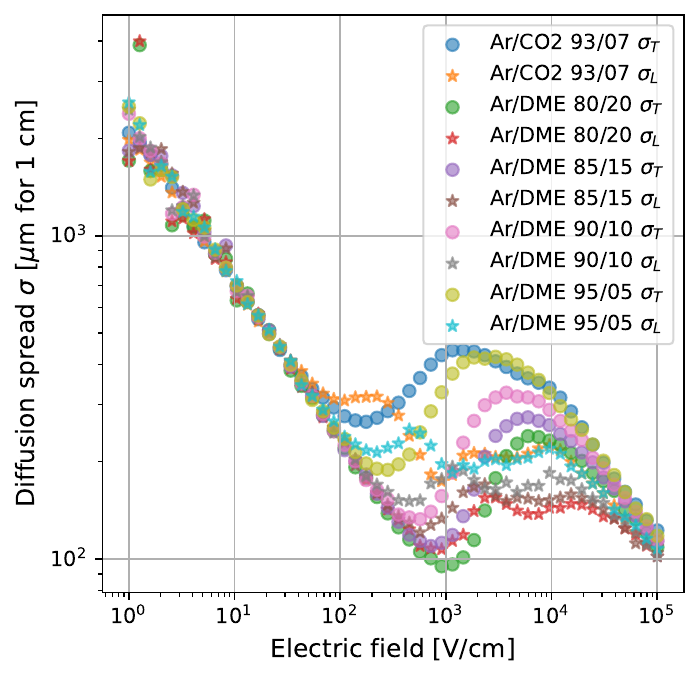}}
\hspace{0.2cm}
\subfigure[\label{fig:drift_ArCO2}]{\includegraphics[scale=0.68]{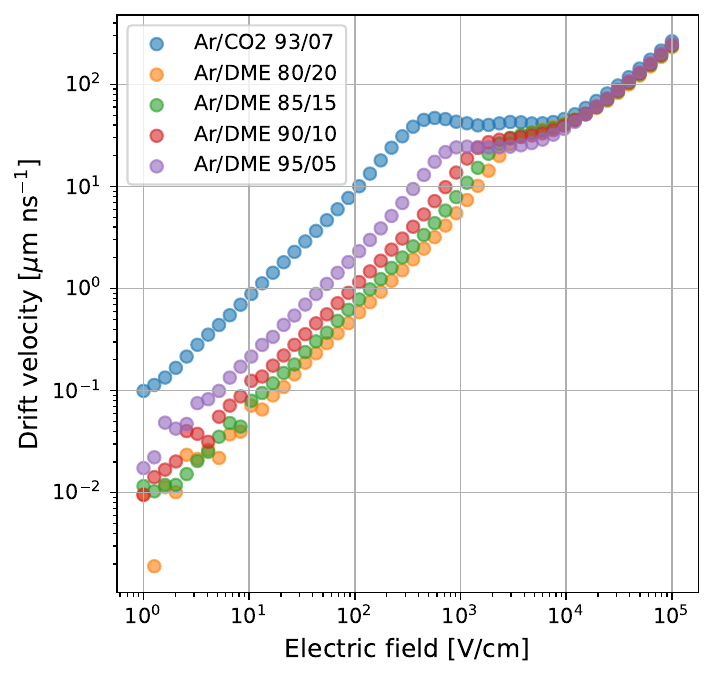}}
\caption{(a) Electron transverse ($\mathrm{\sigma_T}$) and longitudinal ($\mathrm{\sigma_L}$) diffusion spread in $\mu$m after 1 cm drift. (b) Electron drift velocity in cm ns$^{-1}$. The shown diffusion and drift parameters are for various Ar/CO$_2$ and Ar/DME mixtures at 1 atm and 20 $^\circ$C obtained from Garfield++/Magboltz simulations \cite{schindler2024garfield++, biagi1999monte}.}
\label{fig:diffusion_drift_ArCO2}
\end{figure}

The interpreted data contain row-wise information corresponding to a trigger in each pixel. For our purposes, the relevant quantities are pixel coordinates, hit time, and collected charge. These are fully encoded in the \texttt{x}, \texttt{y}, \texttt{TOA}, \texttt{fTOA}, and \texttt{TOT} columns of each hit. Hits are clustered into events based on their temporal proximity, while isolated noise hits within an event are removed using a second-level clustering based on spatial proximity. Basic event selection criteria, including a minimum number of hits per cluster, are applied to improve the signal-to-noise ratio. Using the \texttt{TOA} (25 ns precision) and \texttt{fTOA} (1.5625 ns precision) timestamps together with the drift velocity estimated from Garfield/Magboltz simulations (Fig.~\ref{fig:drift_ArCO2}), the full three-dimensional track can be reconstructed for each event.

\section{MEASUREMENTS WITH PROTOTYPE}
\label{sec:sections}

\begin{figure}
    \centering
    \includegraphics[width=0.6\linewidth]{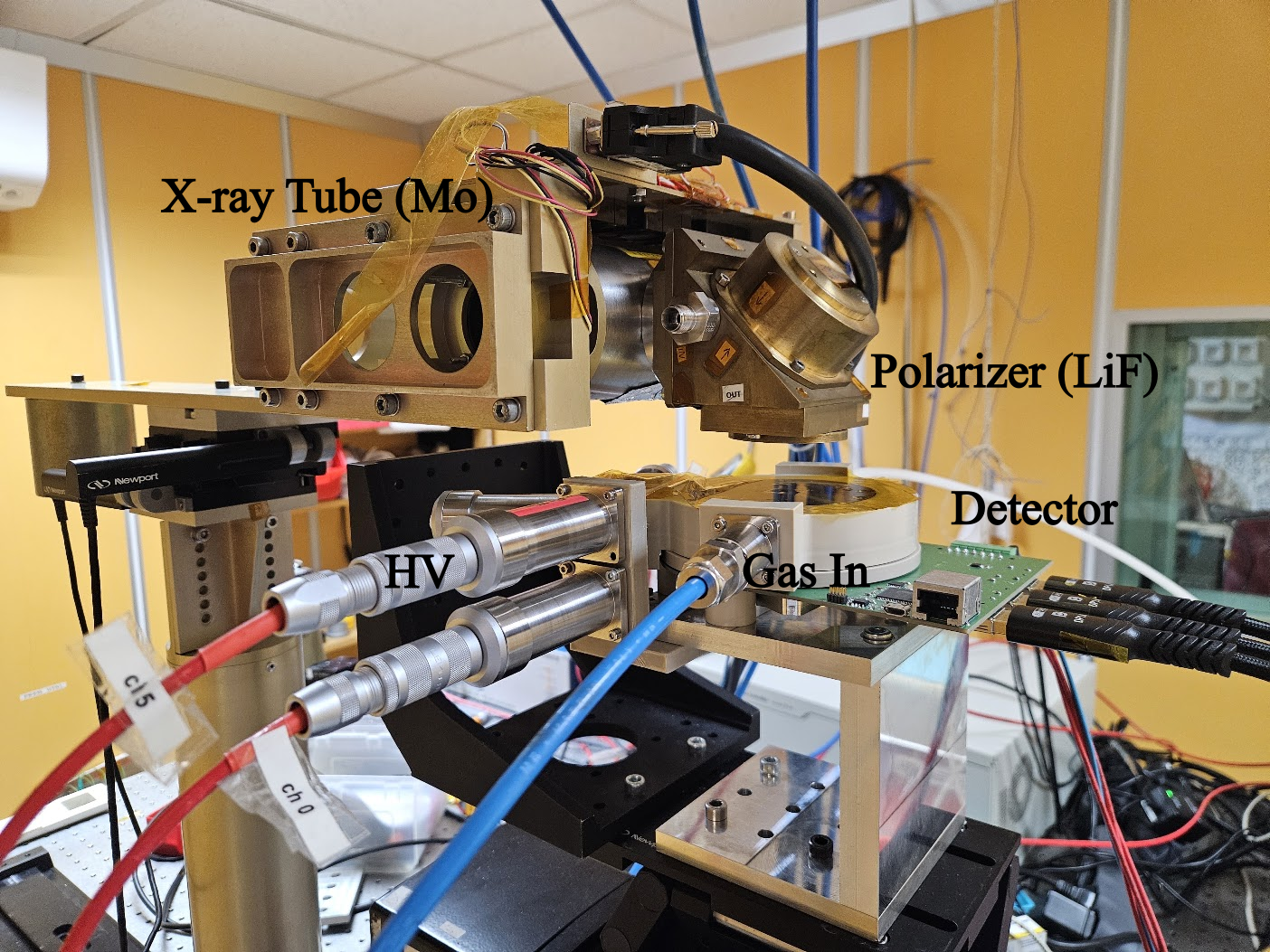}
    \caption{Photograph of the experimental setup. An X-ray tube coupled to a LiF crystal produces polarized 8.7 keV photons via Bragg diffraction, directed into the detector mounted on a rotary/translation stage.}
    \label{fig:experiment_setup}
\end{figure}

For the initial measurements, we assembled the detector prototype with a 1 cm drift gap and 50 $\mu$m beryllium window, and performed the measurements at INAF-IAPS, Rome, in the X-ray calibration facility \cite{muleri_iaps_calib}. The facility has a gas mixing system capable of mixing up to three gas components (Noble, Hydrocarbon, and Electronegative gases). The gas flow was maintained at a pressure of 1.2 bar of Ar/DME 80/20 inside the detector volume. This gas mixture gives diffusion of $\sim$100 $\mu$m for 1 cm of drift and a drift velocity of $\sim$7 $\mu$m ns$^{-1}$ for the electrons. We measured the response of the detector to both unpolarized and polarized X-rays. The experimental setup is shown in Figure~\ref{fig:experiment_setup}.

The output data is a list of pixel triggers and their times ($\S$\ref{sec:RA}). The photoelectron tracks were segregated from this list of pixel triggers by applying time clustering to group the triggers that are close in time ($\Delta t$ = 100 ns). Each segregated track is represented as a set of \texttt{TRK\_SIZE} triggered pixels,
\begin{equation*}
    T = \{ (\texttt{x}_i, \texttt{y}_i, \texttt{TOA}_i, \texttt{TOT}_i) \}_{i=1}^{\texttt{TRK\_SIZE}}
\end{equation*}
where $(\texttt{x}_i, \texttt{y}_i)$ are the spatial coordinates, $\texttt{TOA}_i$ is the time of arrival, and $\texttt{TOT}_i$ is the charge collected (in 40 MHz clock cycles) in the $i$-th pixel. Using a statistical moment-based reconstruction algorithm, the X-ray absorption point (impact point) coordinates [$\texttt{x}_\mathrm{IP}$, $\texttt{y}_\mathrm{IP}$] and the photoelectron ejection direction (azimuthal angle $\Phi_\mathrm{pol}$) were reconstructed \cite{bellazini_moment}.

\begin{figure}[]
\centering
\subfigure[\label{fig:55Fe-spectrum}]{\includegraphics[scale=0.48]{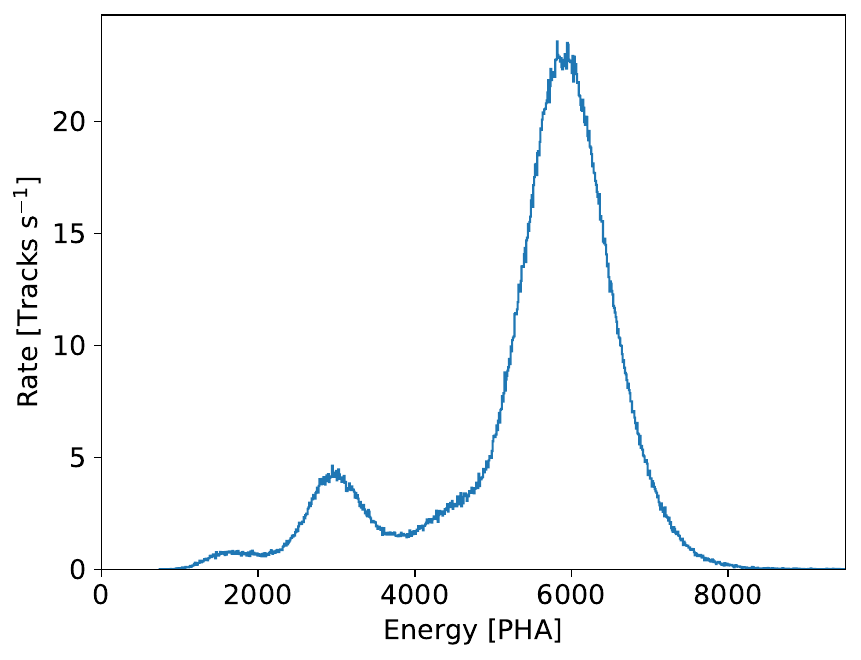}}
\subfigure[\label{fig:55Fe-lcurve}]{\includegraphics[scale=0.36]{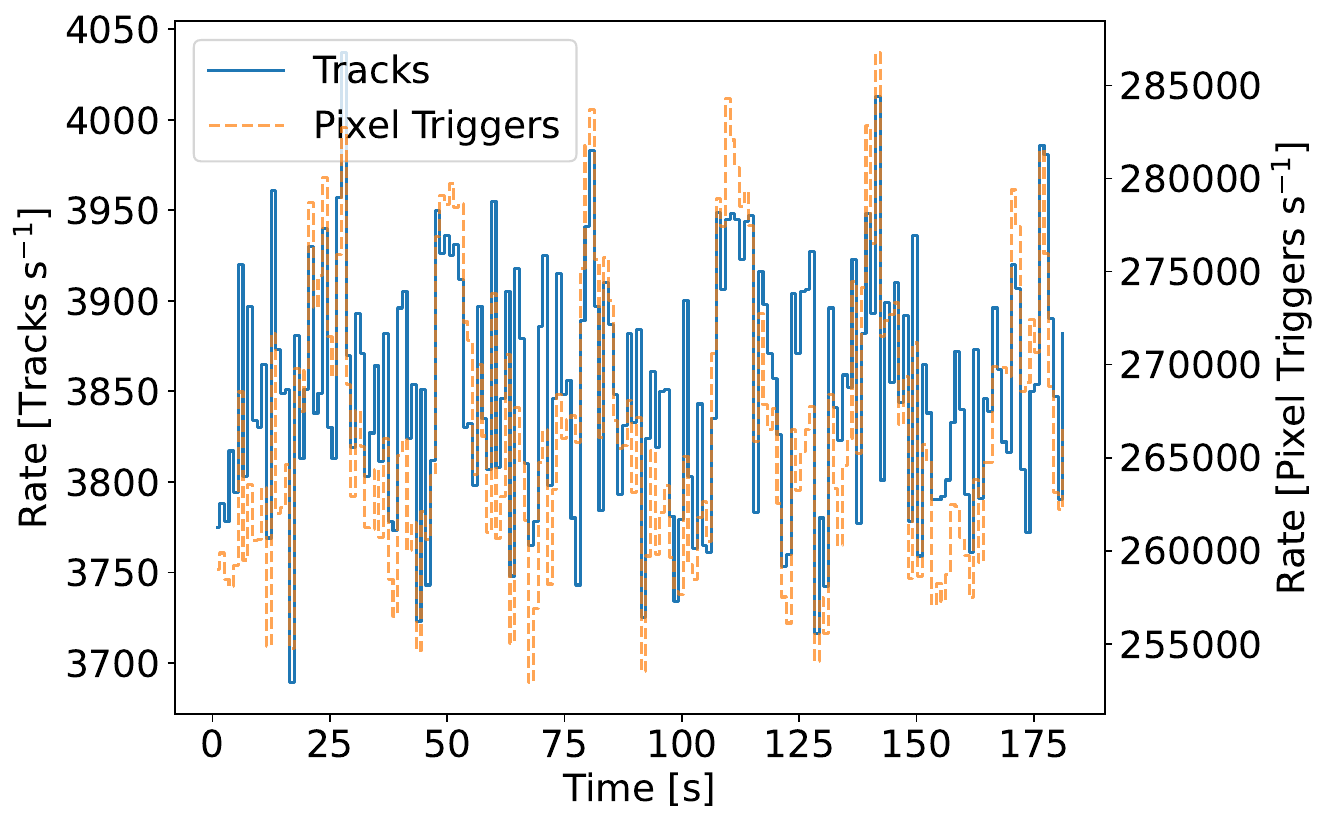}}

\subfigure[\label{fig:55Fe-trk2d}]{\includegraphics[scale=0.41]{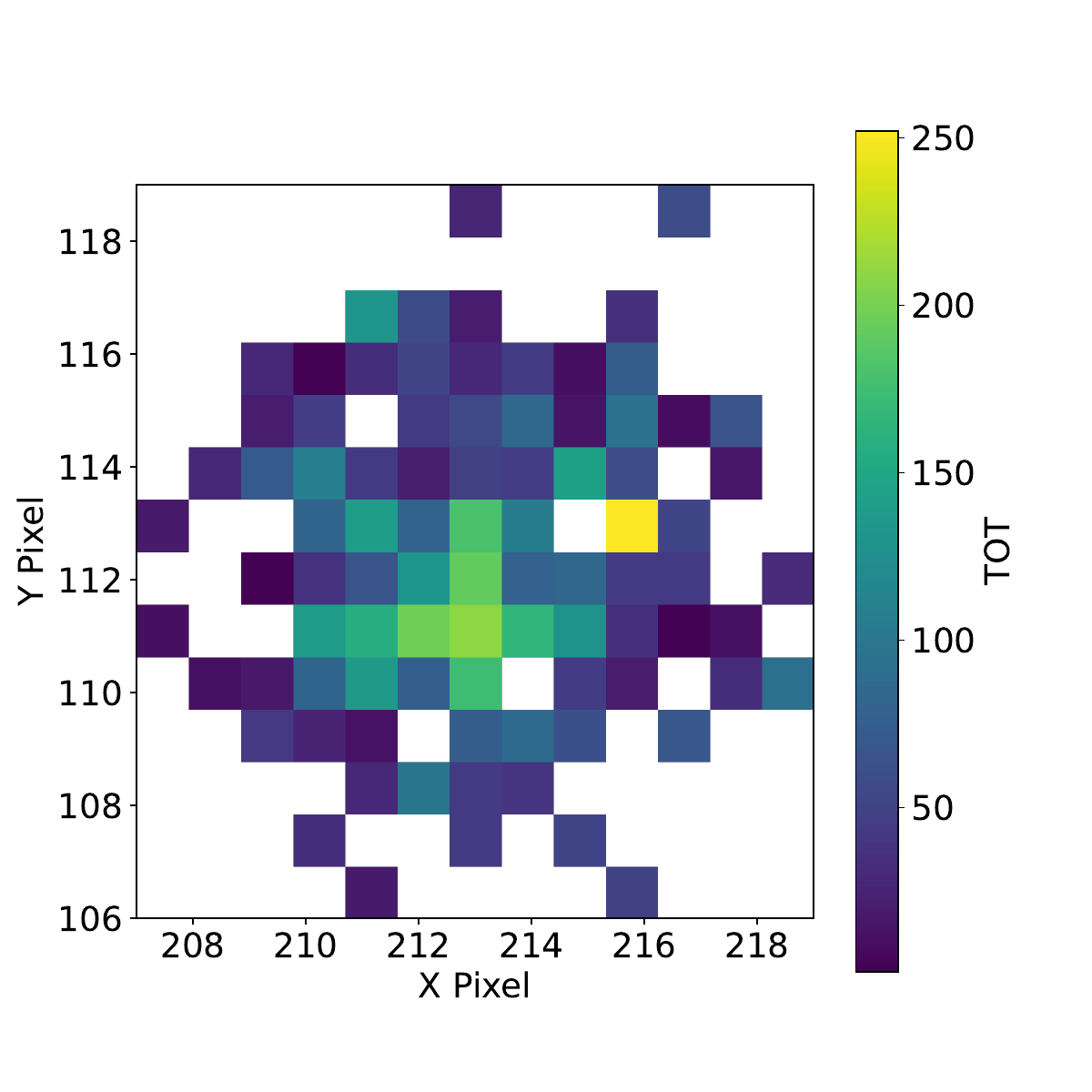}}
\hspace{0.0cm}
\subfigure[\label{fig:55Fe-trk3d}]{\includegraphics[scale=0.41]{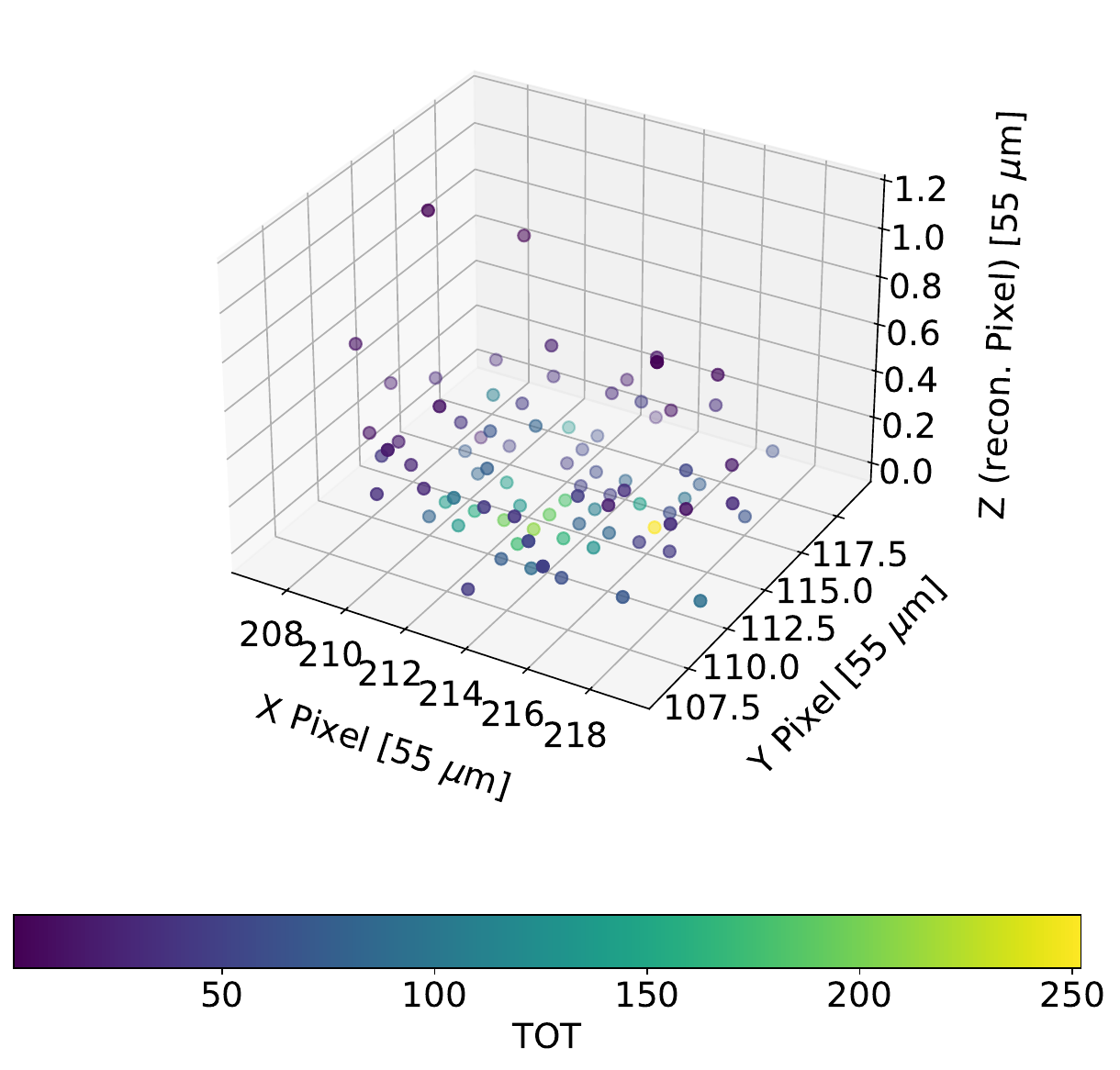}}
\caption{Figure demonstrating the spectroscopic and timing performance of the detector. The active pixel region was illuminated with a $^{55}$Fe radionuclide. \ref{fig:55Fe-spectrum} shows the pixel-averaged charge spectrum; there is the 5.89 keV fluorescence line and 2.7 keV escape peak of Ar. \ref{fig:55Fe-lcurve} The light curve (1 s bin size) recorded over 3 minutes of data acquisition measured a source count rate of $\sim$3800 counts s$^{-1}$. \ref{fig:55Fe-trk2d} is the 2D projection of a sample track and \ref{fig:55Fe-trk3d} is the 3D reconstructed track. The x and y pixels have a resolution of 55 $\mu$m, the pixel pitch of Timpix3, while the z is reconstructed using a 1.5625 ns clock and the drift velocity of $\sim$7 $\mu$m ns$^{-1}$, achieving an equivalent position resolution of $\sim$11 $\mu$m in z.}
\label{fig:diffusion_drift_ArCO2}
\end{figure}

Before performing the polarization measurements, we verified the spectroscopic and timing performance of the detector using the $^{55}$Fe radionuclide. $^{55}$Fe was placed above the beryllium window on a thin Kapton tape $\sim$ 1 cm away from the window. The charge spectrum averaged over the whole active pixel area shows the 5.89 keV emission line and the Ar escape peak (Figure~\ref{fig:55Fe-spectrum}). To verify the timing capability, we reconstructed the track in 3D (Figure~\ref{fig:55Fe-trk3d}). 

\subsection{Unpolarized Flat-field}

A crucial aspect of characterizing any polarimeter is establishing that the detector measures zero modulation (a flat photoelectric angle distribution) for unpolarized sources. Spurious modulation is most prominent at the lower limit of the operational energy band due to larger errors in reconstructing the emission angles of shorter, more rounded tracks (see, for example, Figure~\ref{fig:55Fe-trk2d}). To assess this, the response across the detector's active area must be evaluated using completely unpolarized X-rays. While standard X-ray tubes often produce partially polarized bremsstrahlung, we utilized a Hamamatsu X-ray tube where the electron beam is accelerated parallel to the photon exit window onto a Cu target. In this specific geometry, both the Cu fluorescence (8.05 keV) and the accompanying bremsstrahlung continuum are unpolarized \cite{ratheesh2023polarization}.

\begin{figure}
    \centering
    \includegraphics[width=\linewidth]{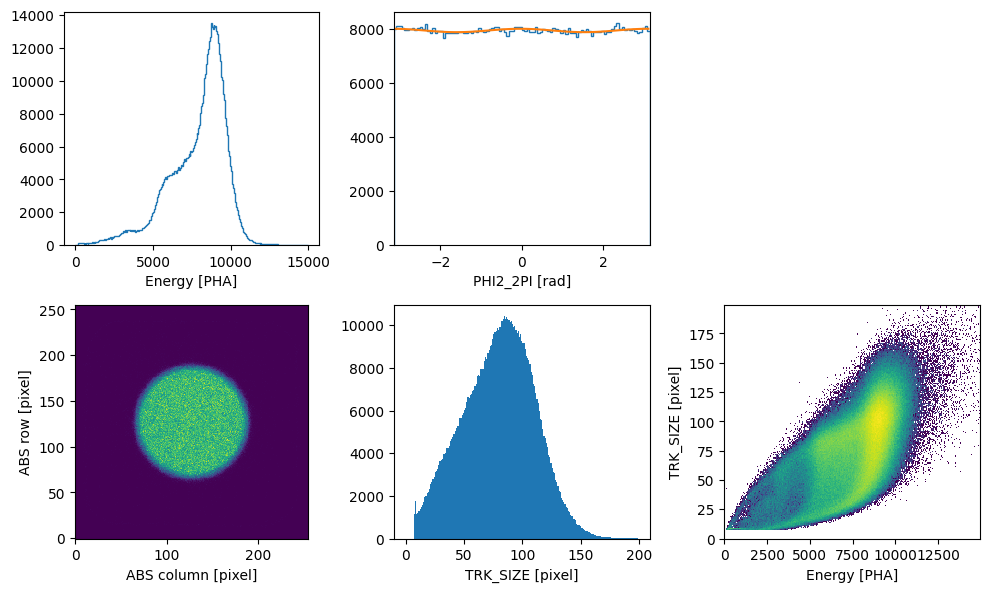}
    \caption{Results of a measurement with an unpolarized 8.05 keV flat field, demonstrating a spurious modulation of $0.8\pm0.2\%$ in the central active pixel area. The bottom panels, from left to right, show the track centroid map within a 60-pixel radius, the histogram of track size (number of pixel triggers per track), and a scatter plot of track energy versus size. The top panels show the charge spectrum of the tracks (left) and the modulation curve (right).}
    \label{fig:unpolarized}
\end{figure}

The resulting spectrum (Figure~\ref{fig:unpolarized}) clearly shows the 8.05 keV Cu fluorescence line alongside the bremsstrahlung continuum and escape peak complex below it. To avoid edge effects caused by drift-field distortions, only interaction points detected within a 60-pixel radius of the detector active area center were considered for the analysis. From the reconstructed tracks, the measured spurious modulation is $0.8\pm0.2\%$. Notably, this low level of spurious modulation is achieved natively, without applying the pixel-to-pixel gain equalization in \textit{IXPE}/GPD \cite{rankin2023equalizing}.

\subsection{Polarized Collimated Beam}

\begin{figure}
    \centering
    \includegraphics[width=\linewidth]{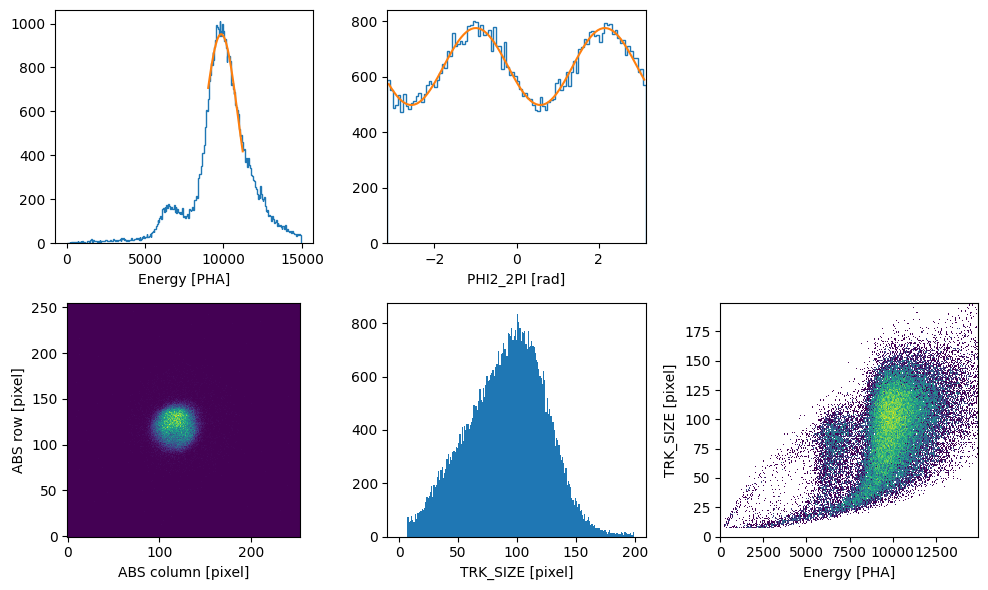}
    \caption{Results of a measurement with a 100\% polarized 8.7 keV collimated beam. For an incident polarization angle set to $-60^\circ$, the measured modulation factor is $21.8 \pm 0.6\%$, with a reconstructed angle of $-56.8 \pm 0.7^\circ$. The bottom panels, from left to right, show the track centroid map, the histogram of track size (number of pixel triggers per track), and a scatter plot of track energy versus size. The top panels show the charge spectrum of the tracks (left) and the modulation curve (right).}
    \label{fig:polarized}
\end{figure}

Measurements with a polarized source were conducted using an Oxford X-ray tube equipped with a molybdenum (Mo) target. The emitted beam was Bragg-diffracted by a LiF 800 crystal at an angle of 45$^\circ$ to produce a nearly 100\% linearly polarized X-ray beam, which was then directed onto the detector through a 2 mm aperture tungsten diaphragm and a 200 $\mu$m thick Cu filter (Figure~\ref{fig:experiment_setup}). Using a rotary stage, the azimuthal polarization angle of the incident beam was set to --60$^\circ$ with respect to the ASIC $x$-axis. The results of the polarized measurement are presented in Figure~\ref{fig:polarized}. The energy spectrum is dominated by the second-order diffraction from the LiF 800 crystal at 8.7 keV and its corresponding escape peak. For this energy, the detector yielded a measured modulation factor of $21.8\pm0.6\%$ and a reconstructed polarization angle of $-56.8\pm0.7^\circ$.

\section{CONCLUSIONS}

GridPix gas detectors with Timepix3 readout ASIC offer a promising technology for the next generation of low- and medium-energy imaging X-ray polarimetry for space-based astronomy. We commissioned a modular X-ray photoelectric polarimeter detector and tested the detector using an Ar/DME 80/20 gas mixture at 1.2 bar on a 1 cm drift gap configuration. We successfully reconstructed the photoelectron track in three dimensions and measured a spectral resolution of ~17\% FWHM at 5.9 keV, demonstrating the timing and spectral capability of the detector. Spurious modulation is strictly contained to $<$ 1\% (at 8.05 keV), notably achieved before applying any pixel-to-pixel gain equalization. Finally, we successfully measured the polarization from 100\%-polarized 8.7 keV photons, yielding a modulation factor of $\sim$22\%. We expect to drive these performance metrics even further by fine-tuning the gas mixture and pressure, applying pixel-to-pixel gain equalization, and adopting a full 3D track reconstruction algorithm.

\acknowledgments 
 
We acknowledge the partial contribution of Progetti di Ricerca di Rilevante Interesse Nazionale del Ministero dell'Università e della Ricerca scientifica (PRIN-MUR) HypeX: High Yield Polarimetry Experiment in X-rays (Hype-X) prot. 2020MZ884C. Parts of this work have received funding from the German Federal Ministry of Education and Research under grant no. 05K22PD1. This work was supported in part by the Italian Ministry of Foreign Affairs and International Cooperation. HM acknowledges the training received at the DRD1 Gaseous Detectors School 2025 and the DRD1 Gaseous Detectors Simulation School 2026. 

\bibliography{report} 
\bibliographystyle{spiebib} 

\end{document}